\documentclass[twocolumn,showpacs,aps,prb,superscriptaddress,floatfix]{revtex4}
\usepackage{graphicx}
\usepackage{dcolumn}
\usepackage{bm}
\begin{document}
\preprint{}

\author{J.P. Clancy}
\affiliation{Department of Physics and Astronomy, McMaster University, Hamilton, Ontario, L8S 4M1, Canada}
\affiliation{Department of Physics, University of Toronto, Toronto, Ontario, M5S 1A7, Canada}

\author{B.D. Gaulin}
\affiliation{Department of Physics and Astronomy, McMaster University, Hamilton, Ontario, L8S 4M1, Canada}
\affiliation{Brockhouse Institute for Materials Research, McMaster University, Hamilton, Ontario, Canada L8S 4M1}
\affiliation{Canadian Institute for Advanced Research, 180 Dundas Street West, Toronto, Ontario, M5G 1Z8, Canada}

\author{A.S. Sefat}
\affiliation{Materials Science and Technology Division, Oak Ridge National Laboratory, Oak Ridge, Tennessee, 37831, USA}

\title{High resolution x-ray scattering studies of structural phase transitions in Ba(Fe$_{1-x}$Cr$_{x}$)$_2$As$_2$}

\begin{abstract}

We have performed high resolution x-ray scattering measurements on single crystal samples of Ba(Fe$_{1-x}$Cr$_x$)$_2$As$_2$ (0 $\le$ x $\le$ 0.335).  These measurements examine the effect of Cr-doping on the high temperature tetragonal ({\it I4/mmm}) to low temperature orthorhombic ({\it Fmmm}) structural phase transition of the parent compound BaFe$_2$As$_2$.  Increasing Cr concentration is found to suppress the structural transition temperature (T$_s$), and reduce the magnitude of the orthorhombic strain ($\delta$).  The doping dependence of the orthorhombic strain, combined with complementary measurements of the high temperature magnetic susceptibility, suggests the presence of a magnetostructural crossover at x $\sim$ 0.05.  In particular, this crossover appears to mark a shift from strong to weak orthorhombicity and from predominantly itinerant to localized magnetic behavior.

\end{abstract}

\maketitle

\section{Introduction}

The magnetic and structural properties of the Fe-based high temperature superconductors have been a subject of intense interest since their discovery in 2008\cite{pnictide_discovery}.  One of the most widely studied families of Fe superconductor are the ``122'' compounds - materials derived from parent compounds with the chemical formula AFe$_2$As$_2$ (A = Ca, Sr, Ba, or Eu).  The 122 compounds have a crystal structure which consists of alternating square FeAs ``active'' layers and A ``blocking'' layers.  The undoped 122 parent compounds undergo concomitant structural and magnetic phase transitions at T$_{s,m}$ = 170 K (A = Ca)\cite{Canfield_Ca_parent}, 203 K (A = Sr)\cite{Tegel_Sr_Eu_parents}, 140 K (A = Ba)\cite{Rotter_PRB_2008} and 190 K (A = Eu)\cite{Tegel_Sr_Eu_parents}.  The structural transition corresponds to a tetragonal ({\it I4/mmm}) to orthorhombic ({\it Fmmm}) structural distortion\cite{Lynn_Physica_C_2009}, while the magnetic transition is associated with the formation of an antiferromagnetic spin density wave (SDW) ground state\cite{Rotter_PRB_2008}.  It has been suggested that orbital physics, and in particular the ordering of Fe 3d$_{xz}$ orbitals, may play an important role in driving this magnetostructural phase transition\cite{Lv_Orbital, Lee_Orbital, Chen_Orbital, Shimojima_Orbital}.

High temperature superconductivity can be induced in the AFe$_2$As$_2$ compounds by doping either the transition metal sites on the active layers or the alkali earth sites on the blocking layers\cite{Canfield_Review, Paglione_Review}.  The series of doped compounds derived from the BaFe$_2$As$_2$ parent material have been particularly well-studied, with superconducting transitions of up to T$_c$ $\sim$ 40 K reported in the case of Ba$_{1-x}$K$_x$Fe$_2$As$_2$\cite{Rotter_K_doped, Avci_K_doped}.  Studies have been performed on Ba(Fe$_{1-x}$TM$_x$)$_2$As$_2$ with TM = Co\cite{Canfield_Review, Sefat_Co_doped, Canfield_Co_Ni_Cu_doped}, Ni\cite{Canfield_Review, Canfield_Co_Ni_Cu_doped, Li_Ni_doped, Ni_Ni_Cu_doped}, Cu\cite{Canfield_Review, Ni_Ni_Cu_doped, Canfield_Co_Ni_Cu_doped}, Rh\cite{Canfield_Review, Ni_Rh_Pd_doped}, Pd\cite{Canfield_Review, Ni_Rh_Pd_doped}, Ir\cite{Wang_Ir_doped}, Pt\cite{Saha_Pt_doped}, Ru\cite{Sharma_Ru_doped, Thaler_Ru_doped, Kim_Ru_doped}, Cr\cite{Sefat_PRB_2009, Budko_PRB_2009, Marty_PRB_2011, Colombier_Cr_doped}, or Mn\cite{Liu_Mn_doped, Kim_Mn_doped, Pandey_Mn_doped}, and Ba$_{1-x}$AE$_x$Fe$_2$As$_2$ with AE = K\cite{Rotter_K_doped, Avci_K_doped, Castellan_K_doped}, Na\cite{Cortes-Gil_Na_doped}.  Both transition metal and alkali metal doping tends to result in the suppression of T$_s$ and T$_m$, which appears to be one of the prerequisites for the development of the superconducting state.  In the case of Co, Ni, Rh, and Pd-doped systems\cite{Canfield_Review, Sefat_Co_doped, Canfield_Co_Ni_Cu_doped, Li_Ni_doped, Ni_Ni_Cu_doped, Ni_Rh_Pd_doped}, doping also leads to a progressive splitting of the structural and magnetic phase transitions, with T$_s$ preceding T$_m$ by as much as 20 K at some concentrations.  Interestingly, superconductivity does not appear to be particularly sensitive to the choice of dopant atom, and has been shown to arise in electron-doped, hole-doped, and even isoelectronically-doped samples.  In fact, it is only for a select group of dopants, such as Cr and Mn (both cases of active layer hole-doping), that superconductivity does not appear to develop at any concentration.

The Cr-doped compound, Ba(Fe$_{1-x}$Cr$_x$)$_2$As$_2$, was the first of the non-superconducting 122 systems to be discovered\cite{Sefat_PRB_2009}.  This system initially generated considerable interest when it was proposed to undergo an unusual tetragonal-to-tetragonal structural phase transition\cite{Sefat_PRB_2009} rather than the tetragonal-to-orthorhombic transition observed in the parent compound and other doped systems.  However, subsequent high resolution x-ray scattering measurements on a Cr-doped sample with x = 0.027 suggest that the structural transition in Ba(Fe$_{1-x}$Cr$_x$)$_2$As$_2$ is in fact a conventional tetragonal-to-orthorhombic transition\cite{Budko_PRB_2009}.  This result is supported by recent neutron scattering measurements\cite{Marty_PRB_2011}, which report Bragg peak intensity changes consistent with a tetragonal-orthorhombic symmetry change at T$_s$.  Studies of the Cr-doped phase diagram reveal a steady suppression of T$_s$ and T$_m$\cite{Sefat_PRB_2009, Marty_PRB_2011}, as in the case of the superconducting 122 systems.  There is still some question as to whether Cr-doping results in a splitting of the structural and magnetic phase transitions, as in Co, Ni, Rh, and Pd-doped BaFe$_2$As$_2$.  While early heat capacity, thermal expansion, and resistivity measurements pointed to a distinct splitting at x = 0.027\cite{Budko_PRB_2009}, neutron scattering results suggest that the structural and magnetic transitions remain concomitant until at least x = 0.335\cite{Marty_PRB_2011}.
 
In this paper, we report detailed x-ray scattering measurements of the structural phase transitions in single crystal samples of Ba(Fe$_{1-x}$Cr$_x$)$_2$As$_2$ (0 $\le$ x $\le$ 0.335).  This work seeks to bridge the gap between the high resolution x-ray scattering work of Bud'ko et al\cite{Budko_PRB_2009}, which is limited to a single doping (x = 0.027), and the x-ray and neutron scattering work of Sefat et al\cite{Sefat_PRB_2009} and Marty et al\cite{Marty_PRB_2011}, which extends across a much wider region of the phase diagram (0 $\le$ x $\le$ 0.47) but with significantly lower experimental resolution.  By comparing our high resolution x-ray scattering measurements with magnetic susceptibility measurements, we can comment on the general properties of the Cr-doped phase diagram as well as the detailed evolution of structural parameters such as the orthorhombic strain.  Our measurements reveal clear evidence of a tetragonal-to-orthorhombic structural transition, with a transition temperature (T$_s$) and orthorhombic order parameter ($\delta$) that decrease significantly as a function of Cr concentration.  Intriguingly, both x-ray and magnetic susceptibility measurements appear to show qualitative changes in behavior at a doping level of x $\sim$ 0.05.  We propose that this concentration represents a magnetostructural crossover that separates a low-doped regime, with strong orthorhombic character and largely itinerant magnetic properties, from a high-doped regime with relatively weak orthorhombicity and well-localized magnetic properties.  These changes can potentially be understood in terms of the effect of chemical disorder on the orbital physics of this system.  These findings provide further evidence of the intimate coupling between structural, magnetic, and orbital degrees of freedom in the 122 family of materials.

\section{Experimental Details}

Single crystal samples of Ba(Fe$_{1-x}$Cr$_x$)$_2$As$_2$ were grown out of a flux mixture of FeAs and CrAs, as previously reported elsewhere\cite{Sefat_PRB_2009}.  Ba(Fe$_{1-x}$Cr$_x$)$_2$As$_2$ exhibits a tendency to form thin plate-like single crystals, with the crystallographic {\it c}-axis perpendicular to the sample surface.  The phase purity of the crystals was characterized using a Scintag XDS 2000 powder x-ray diffractometer, and the unit cell volume was found to increase steadily as a function of Cr-doping.  The dimensions of the samples ranged from $\sim$ 2 $\times$ 2 $\times$ 0.2 mm$^3$ to 6 $\times$ 2.5 $\times$ 0.2 mm$^3$.  The mosaic spread of the crystals was determined from x-ray rocking scans, and was found to vary from 0.06$^{\circ}$ to 0.25$^{\circ}$.

Electron probe microanalysis was performed on the cleaved surface of the Cr-doped single crystals using a Hitachi S3400 Scanning Electron Microscope operating at 20 kV.  The beam current was set to provide $\sim$ 1500 counts/second using a 10 mm$^2$ EDAX detector set for a processing time of 54 $\mu$sec.  The data was reduced using EDAX's Standardless Analysis program.  This analysis indicated that the concentration of Cr present in the crystal structure was less than the concentration of Cr present in the initial solution.  Samples will be denoted by the measured (actual) values of x throughout this manuscript.  The experimental uncertainty in x is $\pm$ 0.01.

Magnetic characterization measurements were performed using a Quantum Design (Magnetic Property Measurement System) SQUID magnetometer.  For a temperature sweep experiment, the samples were cooled to 1.8 K in zero-field (zfc) and data was collected while warming from 1.8 K to 350 K in an applied field of 1 Tesla.  The samples were aligned such that the applied field was parallel to the crystallographic {\it ab}-plane, providing a measure of the in-plane magnetic susceptibility, $\chi_{ab}$.  Details of the {\it c}-axis susceptibility, $\chi_c$, have been reported elsewhere\cite{Sefat_PRB_2009, Budko_PRB_2009}.

\begin{figure}
\includegraphics{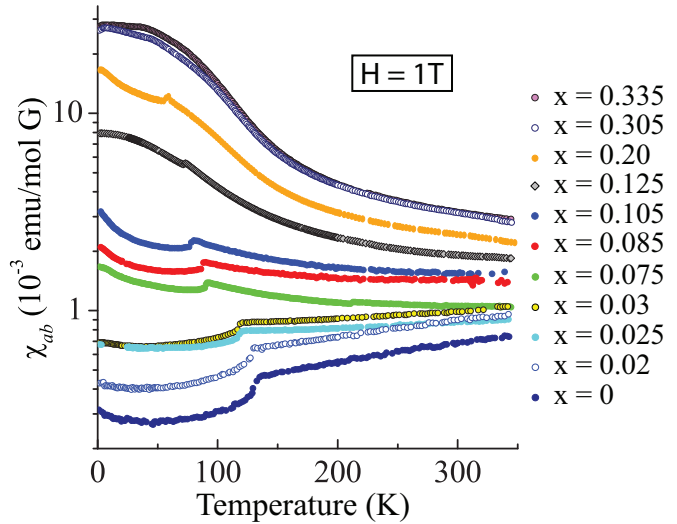}
\caption{(Color online) Temperature dependence of the magnetic susceptibility of single crystal Ba(Fe$_{1-x}$Cr$_x$)$_2$As$_2$ for an applied field of H = 1 T within the {\it ab}-plane.}
\end{figure}

\begin{figure}
\includegraphics{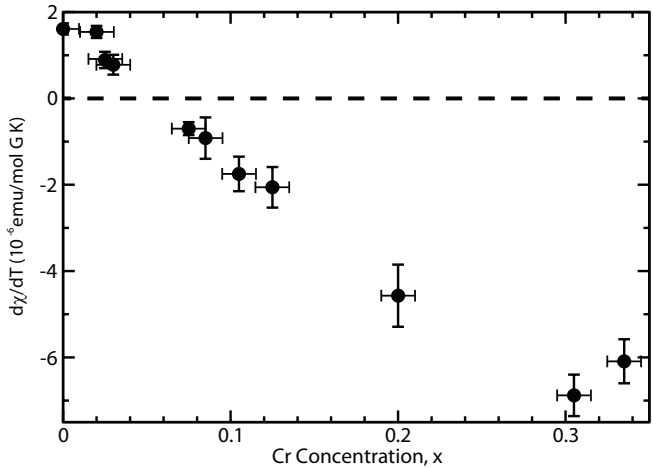}
\caption{(Color online) Doping dependence associated with the slope of the high temperature magnetic susceptibility, d$\chi$/dT, for single crystal Ba(Fe$_{1-x}$Cr$_x$)$_2$As$_2$.  All slopes have been determined from linear fits to $\chi$(T) over a temperature range of 250 K $<$ T $<$ 350 K.}
\end{figure}

X-ray scattering measurements were performed using Cu-K$\alpha_1$ radiation ($\lambda$ = 1.54041 {\AA}) produced by an 18 kW rotating anode x-ray source with a perfect germanium (111) monochromator.  The penetration depth for x-rays of this wavelength is $\sim$ 13 $\mu$m for Ba(Fe$_{1-x}$Cr$_x$)$_2$As$_2$.  Samples were mounted on the coldfinger of a closed-cycle helium cryostat and aligned within a four-circle Huber diffractometer.  The temperature of the sample was maintained to within $\pm$ 0.1 K.  A Bicron NaI scintillation detector was mounted on the detector arm, approximately 64 cm from the sample position.  The angular resolution in this configuration was approximately 0.02$^{\circ}$.  X-ray measurements primarily focused on the (1,1,2) and (1,1,6)$_{HTT}$ Bragg peaks, as indexed using the tetragonal notation of the high temperature (HTT) phase.  Due to the sample morphology, these reflections were studied using transmission geometry.  Additional measurements were also performed on the (0,0,4), (0,0,8) and (1,1,8)$_{HTT}$ Bragg peaks in order to determine the lattice parameters of each sample.

\section{Results and Discussion}

Magnetic susceptibility measurements performed on single crystal samples of Ba(Fe$_{1-x}$Cr$_x$)$_2$As$_2$ (0 $\le$ x $\le$ 0.335) are provided in Figure 1.  As noted in the previous section, these measurements were carried out with a magnetic field of H = 1 T applied within the {\it ab}-plane.  The magnetic susceptibility of Ba(Fe$_{1-x}$Cr$_x$)$_2$As$_2$ is known to be anisotropic\cite{Sefat_PRB_2009, Budko_PRB_2009}, with $\chi_{ab}$ larger than $\chi_c$ at room temperature and showing a much more pronounced anomaly at T$_m$.  Upon Cr-doping, $\chi_{ab}$ undergoes three significant changes.  Firstly, the value of the magnetic transition temperature, T$_m$, steadily decreases as a function of Cr concentration.  The suppression of T$_m$ is evident from the shift in the position of the susceptibility anomaly as x increases.  Secondly, the magnitude of the room temperature susceptibility steadily increases as a function of Cr concentration.  The room temperature value of $\chi_{ab}$ for x = 0.335 is almost 5 times larger than that of the undoped parent compound.  As $\chi_{ab}$ rises, the nature of the susceptibility anomaly also changes, evolving from a sharp, step-like drop (0 $\le$ x $\le$ 0.20), to a more gradual change in slope (x = 0.305, 0.335).  Thirdly, the slope of the high temperature susceptibility, d$\chi_{ab}$/dT, monotonically decreases as a function of Cr concentration.  In fact, the sign of d$\chi_{ab}$/dT changes from positive (x $\le$ 0.03) to negative (x $\ge$ 0.075) as the value of x increases.  This is illustrated by the data in Figure 2, which shows the slope of the high temperature susceptibility as determined from linear fits to $\chi_{ab}$(T) over a temperature range of 250 K to 350 K.  Note that d$\chi_{ab}$/dT appears to drop approximately linearly from x = 0 to x $\sim$ 0.305, suggesting that the crossover in slope should occur at x $\sim$ 0.05.

In previous work by Sefat et al\cite{Sefat_PRB_2009}, the enhancement of the room temperature magnetic susceptibility has been taken as an indication that Cr-doping acts to drive the system towards a ferromagnetic ground state.  This interpretation is supported by a corresponding increase in the Wilson ratio, which relates the susceptibility to the Sommerfeld coefficient, $\gamma$, obtained from specific heat measurements\cite{Sefat_PRB_2009}.  It is also consistent with first principles calculations performed using the linearized augmented plane-wave (LAPW) method\cite{Sefat_PRB_2009}, which predict a ferromagnetic ground state for the intermediate compound BaFeCrAs$_2$ (i.e. at x = 0.50).  However, alternative first principles calculations, performed utilizing the projector augmented wave (PAW) method, arrive at a different conclusion and predict G-type antiferromagnetic order for BaFeCrAs$_2$\cite{Hu_JPCC_2010}.  An antiferromagnetic ground state has also been predicted for the x = 1 end compound, BaCr$_2$As$_2$\cite{Singh_PRB_2009}.  While the magnetic ground state of BaFeCrAs$_2$ has never been measured experimentally, neutron scattering measurements at slightly lower concentrations (0.305 $\le$ x $\le$ 0.47) are consistent with G-type antiferromagnetism\cite{Marty_PRB_2011}.

A consideration of the slope and the detailed form of the high temperature susceptibility, rather than simply the magnitude, suggests an alternative interpretation which may help to reconcile these observations.  In the pure and lightly Cr-doped samples (0 $\le$ x $\le$ 0.03), we observe high temperature magnetic susceptibility with a positive slope and a roughly linear temperature dependence.  This temperature dependence, which cannot be accurately described in either a strictly itinerant (Pauli) or localized (Curie-Weiss) picture, has been attributed to the presence of strong antiferromagnetic correlations above T$_{s,m}$ and the coexistence of both local moments and itinerant electrons\cite{Wang_Susceptibility, Zhang_Susceptibility}.  Within this region of positive d$\chi_{ab}$/dT, the susceptibility clearly cannot be described by a standard Curie-Weiss model.  In contrast, for the highly Cr-doped samples (0.20 $\le$ x $\le$ 0.335) the slope of the high temperature susceptibility is negative, and $\chi_{ab}$(T) can be fit very well by a Curie-Weiss model.  This suggests that the behavior of the highly doped samples is more accurately described in terms of a localized picture, in which the magnetic properties are largely determined by local moments.  The Curie-Weiss constants obtained for these samples yield values of $\theta_{CW}$ = -90 to -170 K, suggesting that the magnetic interactions in this system remain predominantly antiferromagnetic up to x = 0.335, with an energy scale similar to T$_{s,m}$ for the pure compound.  In the intermediate regime (0.075 $\le$ x $\le$ 0.125), the high temperature susceptibility has a negative slope, but the agreement with a Curie-Weiss curve is considerably poorer.  This would appear to suggest a gradual change in magnetic character upon doping, from a region of generally itinerant or mixed itinerant/localized behavior at x $<$ 0.05, to an increasingly well-localized regime at x $>$ 0.05.  This interpretation is supported by the observation of a steady drop in the slope of the resistivity upon doping\cite{Sefat_PRB_2009}, and by the development of antiferromagnetic (rather than ferromagnetic) order at higher Cr concentrations\cite{Marty_PRB_2011}.

Representative x-ray scattering measurements for Ba(Fe$_{1-x}$Cr$_x$)$_2$As$_2$ are provided in Figure 3.  These scans have been performed through the (1,1,2) and (1,1,6)$_{HTT}$ structural Bragg peaks (using the tetragonal notation of the high temperature phase).  For a conventional high temperature tetragonal (HTT) to low temperature orthorhombic (LTO) structural phase transition, as observed in the BaFe$_2$As$_2$ parent compound, one would expect that any Bragg peak which can be indexed as (H,H,L)$_{HTT}$ above T$_s$ should split into two distinct (2H,0,L)$_{LTO}$ and (0,2K,L)$_{LTO}$ peaks below T$_s$.  This form of peak splitting is clearly visible in Figures 3(a) and 3(c) for the case of the x = 0 parent compound.  Figure 3(a) shows a series of representative line-scans through the (1,1,6)$_{HTT}$ peak, collected at temperatures both above and below the structural transition temperature.  Figure 3(c) shows a color contour map composed of many similar scans collected over a wider range of temperatures at 1 K intervals.  The observed peak splitting corresponds to a HTT-LTO structural phase transition at T$_s$ = 135.2 $\pm$ 1.5 K, a value well within the range of reported values for BaFe$_2$As$_2$\cite{Rotter_PRB_2008, Canfield_Review}.  The rapid drop in scattering intensity, the sudden jump in peak position, and the presence of thermal hysteresis all indicate that the structural transition at T$_s$ is discontinuous, or first order, in nature.  This is consistent with previous measurements of the structural phase transitions in pure\cite{Huang_PRL_2008} and Cr-doped\cite{Marty_PRB_2011} BaFe$_2$As$_2$.  As will be discussed in more detail later, the magnitude of the splitting between the (2,0,6)$_{LTO}$ and (0,2,6)$_{LTO}$ Bragg peaks provides a measure of the orthorhombic strain, $\delta = 2(a-b)/(a+b)$, which represents the order parameter for this phase transition.  For the present, it is sufficient to note that the value of $\delta$ = 0.0073 $\pm$ 0.0001 obtained from these measurements is fully consistent with previous x-ray work on BaFe$_2$As$_2$\cite{Rotter_PRB_2008, Canfield_Review}.

\begin{figure}
\includegraphics{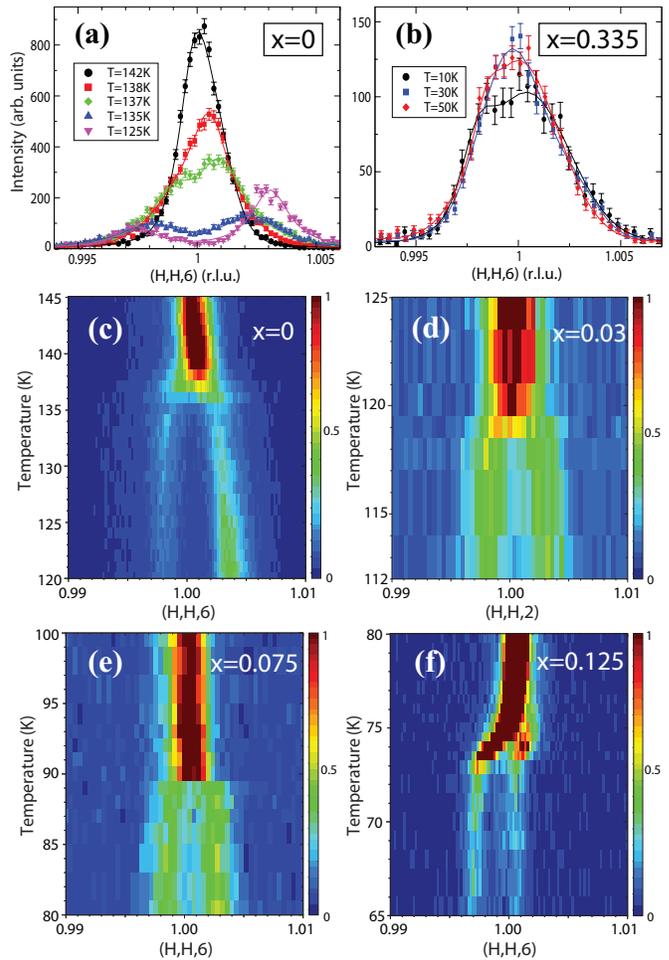}
\caption{(Color online) High resolution x-ray scattering measurements on single crystal Ba(Fe$_{1-x}$Cr$_x$)$_2$As$_2$. Representative line-scans taken through the (1,1,6)$_{HTT}$ Bragg peak are shown for (a) x = 0 (the parent compound) and (b) x = 0.335 as a function of temperature. Color contour maps composed of similar (H,H,6) scans are provided for (c) x = 0, (d) x = 0.03, (e) x = 0.075, and (f) x = 0.125.}
\end{figure}

Figures 3(d),(e), and (f) show similar color contour maps for Cr-doped samples with concentrations of x = 0.03, 0.075 and 0.125, respectively.  As in the case of x = 0, these maps provide evidence of peak splitting consistent with a HTT-LTO structural distortion.  This observation provides confirmation of the high-resolution x-ray measurements previously reported by Bud'ko et al for a sample with x = 0.027\cite{Budko_PRB_2009}.  This peak splitting was likely not observed in the original work by Sefat et al due to the relatively small orthorhombic strain and the considerably lower experimental resolution employed\cite{Sefat_PRB_2009}.  The measurements in Figures 3(d), (e), and (f) show that T$_s$ is gradually suppressed by progressive Cr-doping, with structural transitions occuring at 119.3 $\pm$ 1 K (x = 0.03), 89.5 $\pm$ 0.5 K (x = 0.075), and 73.3 $\pm$ 1.0 K (x = 0.125).  Similarly, the magnitude of the peak splitting, and hence the orthorhombic strain, can be seen to decrease significantly with increasing x.  In the case of the highest doping studied, x = 0.335, the splitting of the orthorhombic peaks is too small to be clearly resolved with the present experimental resolution.  This is illustrated by the representative (H,H,6) scans provided in Figure 3(b).  Despite the absence of distinct peak splitting, there appears to be a slight broadening of the characteristic lineshape which may suggest the presence of a small orthorhombic distortion below T $\sim$ 30 K.  While this broadening is too subtle to allow a reliable determination of T$_s$ (even at base temperature the change in peak width is only $\sim$ 20\%), it does allow us to place a reasonable upper bound on the size of the orthorhombic strain in the sample.  It is also worth noting that the temperature scale associated with this broadening appears to be similar to the magnetic transition temperature for this sample (T$_m$ = 38 $\pm$ 3 K), albeit slightly lower.

The lattice parameters extracted from these x-ray scattering measurements are provided in Figure 4.  These lattice parameters agree very well with previously reported values obtained for powder samples of Ba(Fe$_{1-x}$Cr$_x$)$_2$As$_2$\cite{Sefat_PRB_2009}.  Note that as a result of the symmetry change at T$_s$, the {\it a} and {\it b}-lattice constants are no longer equivalent at base temperature (within the LTO phase).  For convenience we have chosen to describe both the high (T = 295 K) and low (T = 7 K) temperature values of the {\it a} and {\it b} lattice constants in terms of the orthorhombic unit cell ($a_{LTO}$ = $\sqrt{2}$ $\times$ $a_{HTT}$).  As illustrated by the data in Figure 4, the most pronounced effect of Cr-doping is an increase in the inter-layer stacking length, or the crystallographic {\it c}-axis.  The {\it c}-lattice parameter is found to increase approximately linearly with Cr concentration from x = 0 to x = 0.335.  At x = 0.335, the inter-layer stacking distance is $\sim$ 1.6 \% larger than that of the undoped parent compound.  The temperature dependence of the lattice parameters reveals that at base temperature, the {\it c}-lattice constant contracts by $\sim$ 0.5 \%.  In comparison, the {\it a} and {\it b} lattice parameters appear to show much weaker doping dependence.  The {\it a} and {\it b}-axes undergo a small linear increase at lower dopings (x $\le$ 0.125) but remain approximately constant at higher dopings (x $\ge$ 0.125).  This suggests that introducing small concentrations of Cr results in a distortion of the FeAs active layers, while at higher concentrations this effect appears to be washed out by disorder.  Note that an increase in the {\it a} and {\it b} lattice parameters is likely to increase the Fe-As distance that determines orbital overlap.  This could provide a potential explanation for the apparent tendency of Fe electrons to become more strongly localized above x $\sim$ 0.05.  As discussed above, the difference between the {\it a} and {\it b} lattice constants is directly proportional to the magnitude of the peak splitting or the orthorhombic strain.  As the data in Figure 4(a) clearly shows, the difference between {\it a} and {\it b} grows progressively smaller with increasing Cr concentration.  This is consistent with the decrease in orthorhombic strain suggested by Figures 3(c)-(f).  

\begin{figure}
\includegraphics{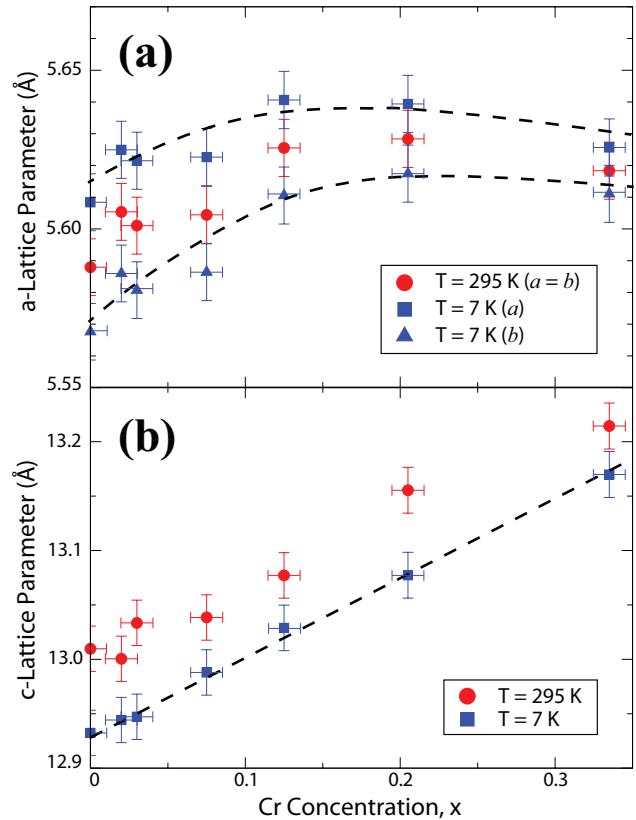}
\caption{(Color online) Temperature and doping dependence of the (a) {\it a} and {\it b}-lattice parameters, and (b) {\it c}-lattice parameter in single crystal Ba(Fe$_{1-x}$Cr$_x$)$_2$As$_2$. Note that the splitting of the {\it a} and {\it b}-lattice parameters for x = 0.335 represents an upper bound determined from the peak broadening observed at T = 7 K. All dashed lines are provided as guides-to-the-eye. }
\end{figure}

By combining the magnetic transition temperatures obtained from our magnetic susceptibility measurements and the structural transition temperatures determined from our x-ray scattering measurements, we can construct an (x,T) phase diagram for Ba(Fe$_{1-x}$Cr$_x$)$_2$As$_2$, as shown in Figure 5.  Here we define the magnetic transition temperatures by the point at which d$\chi$/dT diverges (0 $\le$ x $\le$ 0.20) or changes in slope (x = 0.305, 0.335), and the structural transition temperatures by the point at which a distinct orthorhombic peak splitting can be observed.  Note that on this basis we do not define a structural transition temperature for the x = 0.335 sample, even though there is a possibility that a small orthorhombic distortion may develop in the vicinity of T$_m$.  We observe a gradual suppression of both structural and magnetic transition temperatures with increasing Cr concentration.  The two transitions appear to be coincident across the entire phase diagram, and are not split as in the case of other transition metal dopants such as Co, Ni, Rh, and Pd\cite{Canfield_Review, Sefat_Co_doped, Canfield_Co_Ni_Cu_doped, Li_Ni_doped, Ni_Ni_Cu_doped, Ni_Rh_Pd_doped}.  If such splitting is present in the case of Cr-doping, then it must be less than 2 K in magnitude even at concentrations of up to x = 0.205.  As reported by Sefat et al\cite{Sefat_PRB_2009} and Marty et al\cite{Marty_PRB_2011}, the suppression of T$_s$ and T$_m$ is much more gradual for Cr-doping than for other types of transition metal-doping.  In Ba(Fe$_{1-x}$TM$_x$)$_2$As$_2$ (TM = Co, Ni, Rh, Pd, Cu) T$_s$ and T$_m$ are reduced by a factor of three or more by x = 0.05\cite{Canfield_Review, Sefat_Co_doped, Canfield_Co_Ni_Cu_doped, Li_Ni_doped, Ni_Ni_Cu_doped, Ni_Rh_Pd_doped}.  In Ba(Fe$_{1-x}$Cr$_x$)$_2$As$_2$, this reduction is less than a factor of two.  In addition, the curvature of T$_{s,m}$(x) appears to be more complex than that observed in the superconducting BaFe$_2$As$_2$ compounds (both electron-doped and hole-doped).  For dopings of x $\le$ 0.03, T$_{s,m}$(x) is clearly concave down (d$^2$T$_{s,m}$/dx$^2$ $<$ 0) as in the case of the superconducting 122 systems.  However, for x $\ge$ 0.075, the curvature of T$_{s,m}$(x) appears to reverse in sign (d$^2$T$_{s,m}$/dx$^2$ $<$ 0) and the transition temperatures begin to decrease more slowly.  The presence of persistent magnetism across the Cr-doped phase diagram is believed to be one of the major factors inhibiting the development of superconductivity\cite{Marty_PRB_2011}.  It should be noted that the phase diagram provided in Figure 5 does not extend as far as the neutron phase diagram reported by Marty et al, who measured samples with Cr concentrations as high as x = 0.47\cite{Marty_PRB_2011}.  This means that our phase diagram terminates very close to the border between the SDW and G-type antiferromagnetic phases (x $\sim$ 0.30).  Interestingly, while the region of our phase diagram from 0 $\le$ x $\le$ 0.335 is generally similar to that of Marty et al, the presence of additional samples at lower dopings (x = 0.02, 0.03, and 0.085) reveals more structure to T$_{s,m}$(x) than previously reported.

\begin{figure}
\includegraphics{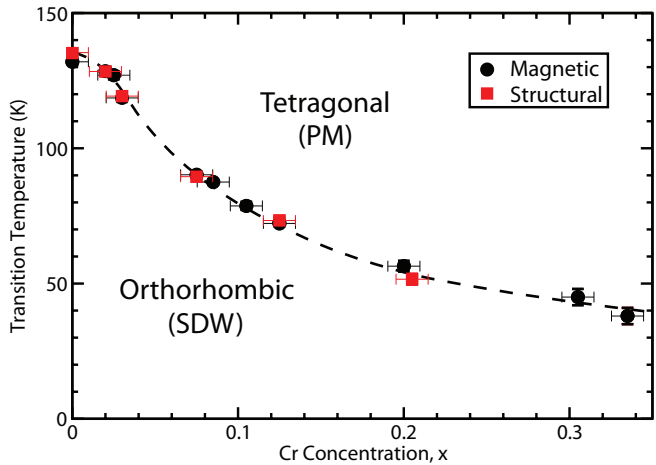}
\caption{(Color online) The (x,T) phase diagram for Ba(Fe$_{1-x}$Cr$_x$)$_2$As$_2$ as determined from magnetic susceptibility and x-ray scattering measurements. Magnetic transitions are indicated for all concentrations, while structural transitions are indicated for x = 0, 0.02, 0.03, 0.075, 0.125, and 0.205. Note that the agreement between the two data sets is such that at some concentrations data points may be overlapping. The dashed line is provided as a guide-to-the-eye.}
\end{figure}

It is instructive to compare this phase diagram to that of the other non-superconducting doped 122 system, Ba(Fe$_{1-x}$Mn$_x$)$_2$As$_2$.  As in the case of Cr-doped BaFe$_2$As$_2$, the Mn-doped series involves hole-doping on the transition metal sites of the active layers.  While detailed measurements of the Mn-doped phase diagram have been somewhat limited by issues related to phase purity and miscibility\cite{Pandey_Mn_doped}, it appears that T$_s$ and T$_m$ are suppressed more rapidly than in the Cr-doped case, but more slowly than in the electron-doped cases\cite{Liu_Mn_doped, Kim_Mn_doped}.  The structural and magnetic transitions in Ba(Fe$_{1-x}$Mn$_x$)$_2$As$_2$ remain coincident for dopings of up to x = 0.102\cite{Kim_Mn_doped}.  Above this concentration the structural transition disappears, and the nature of the magnetically ordered structure changes\cite{Kim_Mn_doped}.  

The Cr-doped phase diagram can also be compared with those of Ba(Fe$_{1-x}$Cu$_x$)$_2$As$_2$ and Ba(Fe$_{1-x}$Ru$_x$)$_2$As$_2$.
The first of these series, Cu-doped BaFe$_2$As$_2$, is an electron-doped compound in which superconductivity is not absent, but is heavily suppressed.  Superconductivity is only observed over a very limited range of dopings (0.035 $<$ x $<$ 0.05), and only at temperatures of $\sim$ 2 K or less\cite{Ni_Ni_Cu_doped}.  However, in spite of the similar suppression of superconductivity, the phase diagram of the Cu-doped system is markedly different from that of Cr-doped BaFe$_2$As$_2$.  Cu-doping results in a much more rapid depression of transition temperatures, and leads to a clear splitting of T$_s$ and T$_m$ even at relatively small dopings (x $<$ 0.025)\cite{Canfield_Review, Canfield_Co_Ni_Cu_doped, Ni_Ni_Cu_doped}.  In fact, with the exception of a much smaller superconducting dome, the Cu-doped phase diagram appears to be almost identical to that of the Co, Ni, Rh, and Pd-doped compounds.  Another useful point of comparison is provided by Ru-doped BaFe$_2$As$_2$, which is an isoelectronically-doped system.  As in the case of Cr-doping, the structural and magnetic transitions in Ba(Fe$_{1-x}$Ru$_x$)$_2$As$_2$ remain coincident across the phase diagram\cite{Sharma_Ru_doped, Thaler_Ru_doped, Kim_Ru_doped}.  However, unlike Ba(Fe$_{1-x}$Cr$_x$)$_2$As$_2$, the Ru-doped series still exhibits superconductivity at concentrations of 0.21 $\le$ x $\le$ 0.36\cite{Sharma_Ru_doped, Thaler_Ru_doped, Kim_Ru_doped}.  These observations serve to show that the suppression of superconductivity is not exclusively linked to the persistence of T$_s$ and T$_m$, just as the occurence of superconductivity is not exclusively linked to the splitting of the structural and magnetic phase transitions.

One of the most interesting parameters that can be extracted from our x-ray data is the orthorhombic strain, $\delta$.  The orthorhombic strain represents the order parameter for the tetragonal-to-orthorhombic structural phase transition at T$_s$, and can be determined from the splitting of the (2H,0,L)$_{LTO}$ and (0,2K,L)$_{LTO}$ Bragg peaks.  The strain is defined in terms of the {\it a} and {\it b} lattice constants, and is given by $\delta = 2(a-b)/(a+b)$.  The doping dependence of the orthorhombic strain is illustrated in Figure 6.  Figure 6(a) shows a series of representative (H,H,6) scans performed for samples with Cr concentrations of x = 0, 0.03, 0.075, 0.205, and 0.335.  These scans were collected at base temperature (T = 7 K), where the orthorhombic strain is assumed to have fully developed and reached its maximal value.  The line scans were fit to a two Lorentzian lineshape, and the peak centers extracted from these fits were used to obtain the values of $\delta$(x).  The doping dependence of $\delta$ is illustrated in Figure 6(b).  Note that $\delta$(x) is approximately constant at lower dopings (x $\le$ 0.03), and then begins to decrease linearly at higher dopings (x $\ge$ 0.075).  To provide a comparison with the previously reported data in the literature\cite{Budko_PRB_2009}, the orthorhombic strain reported by Bud'ko et al for x = 0.027 has also been included in this plot (as denoted by the red square).  The x = 0.027 data point clearly falls on the same trend as the data from the present study.  By extrapolating the linear trend in $\delta$(x) at higher dopings, we can predict that the orthorhombic strain should vanish at x = 0.50 $\pm$ 0.05 (assuming this trend remains valid above x = 0.335).

\begin{figure}
\includegraphics{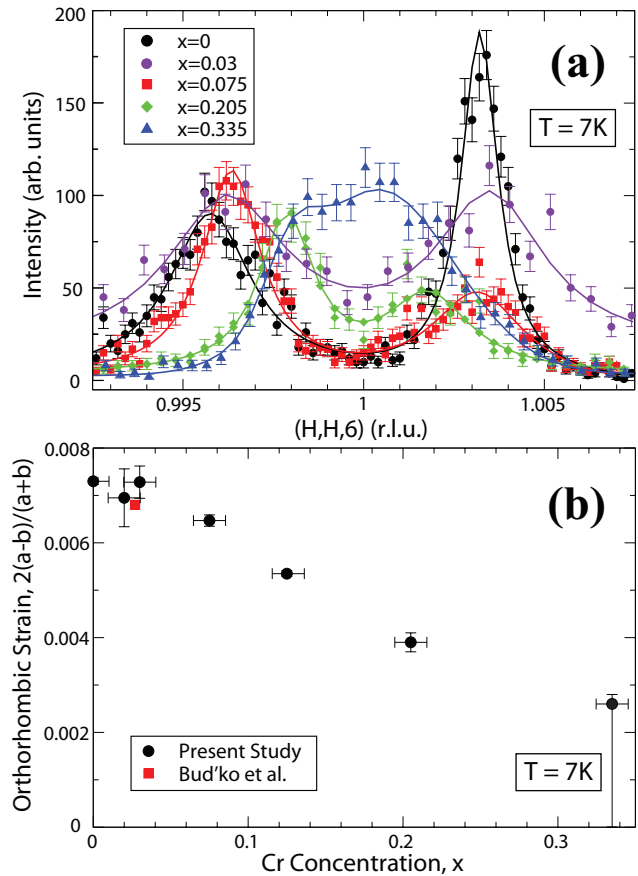}
\caption{(Color online) Doping dependence of the orthorhombic strain, 2(a-b)/(a+b), in Ba(Fe$_{1-x}$Cr$_x$)$_2$As$_2$. (a) Representative (H,H,6) scans collected at base temperature (T = 7 K) for samples with x = 0, 0.03, 0.075, 0.205, and 0.335. (b) Orthorhombic strain as a function of Cr concentration at base temperature.  The data point at x = 0.335 represents an upper bound based on the two peak fit provided in (a).}
\end{figure}

One can remove any explicit dependence on the nominal/actual doping levels of these data sets by plotting the structural transition temperature as a function of orthorhombic strain, as shown in Figure 7(a).  Both T$_s$ and $\delta$ can be extracted directly from the x-ray scattering data shown in Figure 3.  In the case of x = 0.335, where two clearly distinguishable peaks cannot be observed, we can still determine a useful upper bound for T$_s$ and $\delta$ based on the observed magnetic transition temperature (T$_m$ = 38 $\pm$ 3 K) and the results of a two peak fit to the base temperature data set (as shown in Figure 6(a)).  This plot shows that there is a clear linear relationship between T$_s$ and $\delta$ at low orthorhombic strains (i.e. for x $\ge$ 0.075).  Note that the line-of-best-fit through this linear regime extrapolates directly through the origin without any artificial constraint.  This demonstrates that the structural transition temperature is directly proportional to the magnitude of the orthorhombic strain at higher dopings, and that the two quantities tend continuously to zero together.  At higher orthorhombic strains (i.e. for x $\le$ 0.03) the data points begin to deviate from this linear trend.  These deviations are not a particularly subtle effect, and the discrepancies between the data points and the line of best fit can be up to 10 or 20 K in size.  In addition, these deviations occur in the range of dopings where the structural transitions are the most clearly defined, and the error bars in T$_s$ are among the smallest.

\begin{figure}
\includegraphics{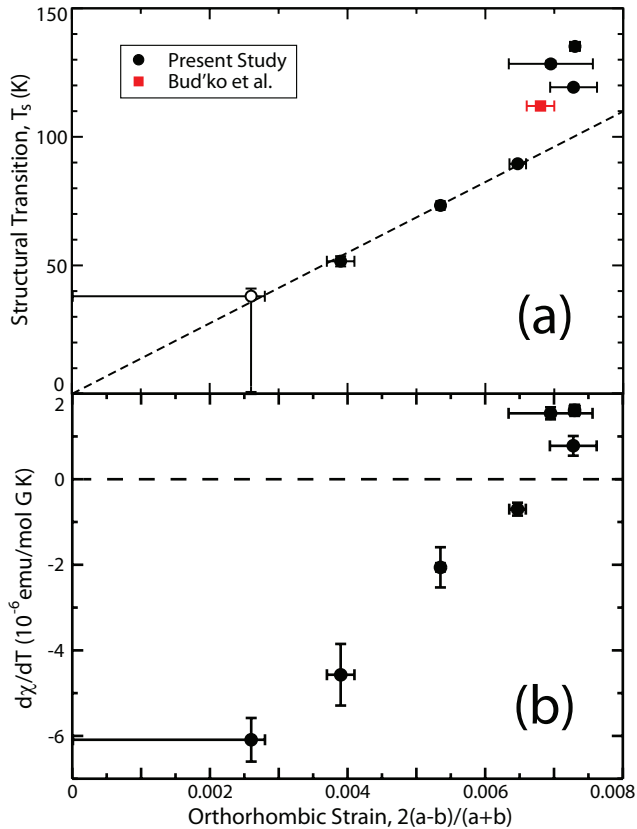}
\caption{(Color online) (a) Observed structural transition temperatures in Ba(Fe$_{1-x}$Cr$_x$)$_2$As$_2$ plotted as a function of orthorhombic strain, 2(a-b)/(a+b). The dashed line represents a fit to the data for x $\ge$ 0.075. The open data point represents an upper bound for x = 0.335 based on the observed magnetic transition temperature and a two peak fit to the base temperature data set. (b) The slope of the high temperature magnetic susceptibility, d$\chi$/dT, as a function of orthorhombic strain.}
\end{figure}

It is important to note that the qualitative changes observed between x = 0.03 and x = 0.075 are reflected in both x-ray scattering and magnetic susceptibility measurements.  This is emphasized by Figure 7(b), which relates the slope of the high temperature magnetic susceptibility to the magnitude of the orthorhombic strain.  It is evident from a comparison of Figures 7(a) and 7(b) that the distinction between the high-doped and low-doped regimes is also marked by the point at which the slope of $\chi$(T) changes from negative to positive.  In fact, the jump from x =0.03 to 0.075 marks a crossover in the doping dependence of the orthorhombic strain, a sign change in the slope of the high temperature magnetic susceptibility, and an inflection point for the structural and magnetic transition temperatures, T$_{s,m}$(x).  These results demonstrate that there is a strong coupling between the structural and magnetic properties of Ba(Fe$_{1-x}$Cr$_x$)$_2$As$_2$, and that x $\sim$ 0.05 represents some form of magnetostructural crossover.  This concentration bears an intriguing similarity to the optimal dopings reported for the superconducting Co, Ni, Rh, and Pd-doped BaFe$_2$As$_2$ compounds, which range from approximately x = 0.04 to x = 0.07\cite{Canfield_Review, Sefat_Co_doped, Canfield_Co_Ni_Cu_doped, Li_Ni_doped, Ni_Ni_Cu_doped, Ni_Rh_Pd_doped}.  Recent neutron scattering measurements on Ba(Fe$_{1-x}$Cr$_x$)$_2$As$_2$ have revealed the presence of a doping-induced magnetic transition at x $\sim$ 0.30, which is associated with a rapid drop in the size of the ordered magnetic moment and the development of a G-type antiferromagnetic ground state\cite{Marty_PRB_2011}.  The changes we observe at x $\sim$ 0.05 are more subtle in nature, and almost certainly arise from different physical origins.

We propose that the importance of Fe 3d orbital physics may provide a potential explanation for the crossover at x $\sim$ 0.05.  The tetragonal-to-orthorhombic symmetry change observed at T$_s$ is believed to be driven by ordering of Fe 3d$_{xz}$ electronic orbitals\cite{Lv_Orbital, Lee_Orbital, Chen_Orbital, Shimojima_Orbital}.  As a result, both the orthorhombic strain (which is related to the magnitude of the orthorhombic distortion) and the in-plane lattice constants (which are related to the bond angles and distances within the FeAs active layers) can provide potential insight into the orbital physics of Ba(Fe$_{1-x}$Cr$_x$)$_2$As$_2$.  In the lightly-doped regime, we observe a large, constant, orthorhombic strain, and in-plane lattice constants which systematically increase as a function of Cr concentration.  These results indicate a well-defined orthorhombic distortion, consistent with the presence of orbital order.  In this region of the phase diagram T$_s$ and T$_m$ decrease quite rapidly and the high temperature magnetic susceptibility is indicative of strong antiferromagnetic interactions with predominantly itinerant or mixed itinerant/localized magnetic character.  In the higher-doped regime, the orthorhombic strain begins to steadily decrease, and the magnitude of the {\it a} and {\it b} lattice constants becomes almost independent of concentration.  This indicates significant, and progressively larger, reductions in the size of the orthorhombic distortion, and suggests a gradual suppression or disruption of orbital order.  This region of the phase diagram is also marked by more gradual suppression of T$_s$ and T$_m$ and an increasing tendency towards localized magnetic moments.  We can interpret these observations in terms of: (i) a lightly-doped regime with structural and magnetic properties largely dependent on orbital physics (as in the parent compound and other doped 122 systems), and (ii) a heavily-doped regime in which orbital effects are increasingly disrupted by the presence of chemical disorder.

\section{Conclusions}

In summary, we have performed detailed x-ray scattering measurements on single crystal samples of non-superconducting, hole-doped Ba(Fe$_{2-x}$Cr$_x$)$_2$As$_2$ (0 $\le$ x $\le$ 0.335).  This study fills an important niche between previous high-resolution single-doping measurements and recent lower-resolution multiple-doping measurements.  Our measurements provide clear evidence of the peak splitting associated with tetragonal-orthorhombic structural phase transitions, in agreement with the high resolution x-ray work reported by Bud'ko et al on Ba(Fe$_{0.973}$Cr$_{0.027}$)$_2$As$_2$\cite{Budko_PRB_2009}.  We observe a gradual depression of the structural and magnetic phase transitions with increasing Cr concentration, and find no apparent separation between T$_s$ and T$_m$ for dopings of up to x = 0.205.

We have performed systematic measurements of the orthorhombic strain, $\delta = 2(a-b)/(a+b)$, across the Cr-doped phase diagram.  Our measurements indicate that the magnitude of the strain is approximately constant at low dopings (x $\le$ 0.03), and decreases linearly at higher dopings (x $\ge$ 0.075).  The crossover between these two regimes also coincides with the point at which the slope of the high temperature magnetic susceptibility, d$\chi$/dT, changes from positive to negative.  This suggests that the crossover at x $\sim$ 0.05 is associated with underlying changes in both the structural and magnetic properties of Cr-doped BaFe$_2$As$_2$, marking a shift from strong to weak orthorhombicity and from predominantly itinerant to localized magnetic character.  This crossover is distinctly different from the magnetic transition reported at x $\sim$ 0.30\cite{Marty_PRB_2011}, whereupon the SDW ground state is replaced by G-type antiferromagnetism and the average ordered moment is dramatically reduced.  We also note that this magnetostructural crossover occurs at a Cr concentration which closely matches the optimal dopings reported for superconducting 122 compounds such as Co, Ni, Rh, and Pd-doped BaFe$_2$As$_2$ (x = 0.04 to 0.07)\cite{Canfield_Review, Sefat_Co_doped, Canfield_Co_Ni_Cu_doped, Li_Ni_doped, Ni_Ni_Cu_doped, Ni_Rh_Pd_doped}.  We hope these results will help to shed further light on the complex interplay between structure, magnetism, orbital order and superconductivity in the Fe-based superconductors.

\begin{acknowledgments}

The authors would like to acknowledge L.H. VanBebber for assistance with sample preparation and J.J. Wagman for assistance with x-ray scattering measurements.  We would also like to thank P.C. Canfield for helpful comments and Y. Mozharivskyj for useful discussion.  Research at McMaster University was supported by NSERC of Canada.  Research at ORNL was supported by the U.S. Department of Energy, Office of Basic Energy Sciences, Materials Sciences and Engineering Division.

\end{acknowledgments}

\end{document}